\documentclass[10pt,journal,epsfig]{IEEEtran}
\usepackage{amsmath}
\usepackage{amssymb}
\usepackage{psfrag}

\usepackage{graphicx}
\usepackage{epstopdf}
\usepackage{todonotes}
\usepackage{algorithm}
\usepackage{algorithmic}
\usepackage{framed}
\usepackage{comment}
\usepackage{subeqnarray}
\usepackage{cite}
\usepackage{color}
\usepackage{changebar}
\ifCLASSOPTIONcompsoc
 \usepackage[caption=false,font=normalsize,labelfont=sf,textfont=sf]{subfig}
\else
 \usepackage[caption=false,font=footnotesize]{subfig}
\fi

\newtheorem{theorem}{Theorem}
\newtheorem{lemma}{Lemma}

\newtheorem{remark}{Remark}
\newtheorem{proposition}{Proposition}
\begin{document}
\title{Joint Power and Trajectory Design for Physical-Layer Secrecy in the UAV-Aided Mobile Relaying System}
\author{Qian Wang, Zhi Chen, \IEEEmembership{Senior Member,~IEEE}, and Shaoqian Li, \IEEEmembership{Fellow,~IEEE}
\thanks{The authors are with the National Key Laboratory of Science and Technology on Communications, University of Electronic Science and Technology of China, Chengdu 611731, China (e-mail: wangqianfresh@hotmail.com, chenzhi@uestc.edu.cn, lsq@uestc.edu.cn).}}
\maketitle
\begin{abstract}
 Mobile relaying is emerged as a promising technique to assist wireless communication, driven by the rapid development of unmanned aerial vehicles (UAVs). In this paper, we study secure transmission in a four-node (source, destination, mobile relay, and eavesdropper) system, wherein we focus on maximizing the secrecy rate via jointly optimizing the relay trajectory and the source/relay transmit power. Nevertheless, due to the coupling of the trajectory designing and the power allocating, the secrecy rate maximization (SRM) problem is intractable to solve. Accordingly, we propose an alternating optimization (AO) approach, wherein the trajectory designing and the power allocating are tackled in an alternating manner. Unfortunately, the trajectory designing is a nonconvex problem, and thus is still hard to solve. To circumvent the nonconvexity, we exploit sequential convex programming (SCP) to derive an iterative algorithm, which is proven to converge to a Karush-Kuhn-Tucker (KKT) point of the trajectory design problem. The simulation results demonstrate the efficacy of the joint power and trajectory design in improving the secrecy throughput.

\end{abstract}
\begin{IEEEkeywords}
UAV, Mobile relay, Physical-layer security, Joint power and trajectory design
\end{IEEEkeywords}

\section{Introduction}
 \emph{Physical-layer security}, employing the fundamental characteristics of the transmission medium to provide secure communication, has become a hot research area recently.
 Aspects of secrecy at the physical layer has been studied for decades since Wyner's seminal work \cite{ADWyner1975}. Corresponding studies have been further extended to wireless
cooperative networks, among which cooperative relaying is exploited as a popularized technique to enhance wireless physical-layer security; see e.g., \cite{QL2015,LDZH2010}.

Nevertheless, conventional relay-assisted strategies assumed a \emph{static relaying} model, i.e., the relays' deployment locations are fixed.
Motivated by the continuous development of unmanned aerial vehicles (UAVs), there is growing interest in extending the cooperative relaying to the case employing \emph{UAV-aided mobile relaying} \cite{PZKY,YZRZ2016,QW}.
Our previous work \cite{QW} is the first to establish the utility of UAV-aided mobile relaying in facilitating the security of wireless systems, in which,
 however, the UAV trajectory design has not been considered.
 This motivates our present work which addresses the joint optimization of the source/relay transmit power and the relay trajectory in maximizing the secrecy rate.

 This paper focuses on a UAV-aided mobile relaying system consisting of a source, a destination, a mobile relay, and an eavesdropper.
 We assume that the mobile relay has a buffer of sufficiently large size. Specifically, our goal is to jointly optimize the relay trajectory and the source/relay power allocations
for maximizing the secrecy rate, while satisfying the practical mobility and \emph{information-causality} constraints. The mobility constraints include the relay's initial and final location constraints as well as the speed constraint. The information-causality constraint guarantees that the relay can only forward
the data that has already been decoded. The resulting secrecy rate maximization (SRM) problem is challenging to solve, due to the coupled design aspects, i.e., the trajectory designing and the power allocating. Our main contributions are summarized below.
\begin{enumerate}
\item 
We exploit the alternating optimization (AO) method to tackle the SRM problem, which is
suboptimal but easy-to-implement. Its main idea is to alternatingly tackle the SRM problem with given trajectory, i.e., the \emph{power allocation} problem and the SRM problem with fixed power allocation, i.e., the \emph{trajectory optimization} problem.

\item However, the trajectory optimization problem is a nonconvex problem, which is still hard to solve. To handle this, we derive successive convex approximate subproblems of the trajectory optimization problem, via which we develop a sequential convex programming (SCP)-based algorithm. In addition, we prove that the SCP-based algorithm has Karush-Kuhn-Tucker (KKT) point convergence guarantee.

\end{enumerate}
\section{System Model and Problem Statement}\label{sec:system-model}
Consider a Gaussian wiretap channel scenario including a source (\emph{Alice}), a destination (\emph{Bob}), an eavesdropper (\emph{Eve}) and a mobile relay. All the terminals are equipped with a single antenna. The direct links from Alice to Bob and to Eve are assumed to be negligible, because of e.g., severe blockage (mountains and/or buildings). The task of relaying the confidential message to Bob is assigned to the UAV-based mobile relay, via which air-to-ground channels can be established. We assume that the mobile relay has a buffer of sufficiently large size and operates in a frequency division duplex (FDD) mode, allocating equal bandwidth for data transmission and reception.

 Suppose that a UAV flying at a fixed altitude $H$ is employed as a mobile relay for a finite time horizon $T$. Practically, $H$ may correspond to the minimum altitude where the UAV does not have to ascend or descend frequently so as to avoid mountains or buildings.
 The time horizon $T$ is discretized into $N$ equally spaced time slots, i.e., $T=N\delta_t$, with $\delta_t$ representing the slot length. To guarantee that the UAV's location can be assumed to be approximately constant within each slot, the slot length $\delta_t$ is chosen to be sufficiently small. Without loss of generality, all communication nodes are placed in the three-dimensional (3D) Cartesian coordinate system, with Alice, Eve and Bob located on the ground with coordinates $(0,0,0)$, $(E,S,0)$ and $(D,0,0)$, respectively, while the UAV has time-varying coordinate $(x[n],y[n],H), n=1,...,N$, with $x[n]$ and $y[n]$ denoting the UAV's x- and y-coordinates, respectively. In view of practicality, the initial and final UAV coordinates depend on factors like the launching/landing coordinates and pre-mission/post-mission trajectories, etc. Therefore,
we consider the scenario where the UAV has predetermined initial and final coordinates denoted as $(x_0,y_0,H)$ and $(x_F,y_F,H)$, respectively. As mentioned before, the slot length is sufficiently small so that the UAV flies with an approximately constant velocity during each slot. Consequently, the UAV speed can be calculated as $\frac{\sqrt{(x[n+1]-x[n])^2+(y[n+1]-y[n])^2}}{\delta_t}, n=1,...,N-1$. Suppose that the UAV's maximum speed is $\tilde V$. Then the relay has the mobility constraints, where both its initial/final locations constraints and speed constraint are included, shown as
\begin{subequations}
\begin{align}
(x[1]-x_0)^2+(y[1]-y_0)^2\le V^2,&\label{va}\\
(x[n+1]-x[n])^2+(y[n+1]-y[n])^2\le V^2,&\nonumber\\
 n=1,...,N-1,&\label{vb}\\
(x_F-x[N])^2+(y_F-y[N])^2\le V^2,&\label{vc}
\end{align}
\end{subequations}
where $V=\tilde V\delta_t$ is the maximum distance the UAV can move within a single slot.

Let $h_{ar}[n]$, $h_{re}[n]$, and $h_{rd}[n]$ denote the channel gains of the Alice-UAV link, the UAV-Eve link, and the UAV-Bob link at the $n$th time slot, respectively.
Analogous to \cite{YZRZ2016}, we consider a free-space path loss model:
\begin{align}
h_{ij}[n]=\beta_0d{_{ij}^{-2}}[n], ij\in\{ar,rd,re\}, n=1,...,N,
\label{channel-model1}
\end{align}
where
$$d_{ar}[n]=\sqrt{H^2+x^2[n]+y^2[n]},$$
$$d_{rd}[n]=\sqrt{H^2+(D-x[n])^2+y^2[n]},$$
 and $$d_{re}[n]=\sqrt{H^2+(E-x[n])^2+(S-y[n])^2}$$
  are the distances of the Alice-UAV link, the UAV-Bob link,
and the UAV-Eve link at slot $n$, respectively, and $\beta_0$ represents the reference channel power gain at distance $d_0=1$ meter. Now, define $\gamma_{ij}[n]\triangleq h_{ij}[n]/\delta^2, ij\in\{ar,rd,re\}, n=1,...,N$,
where $\delta{^2}$ denotes the noise power.
For simplicity, we focus on the scenario that global channel state information (CSI) is available, including the CSI for Eve. This is possible in situations where Eve is an unauthorized user as far as the information for Bob is concerned, while it is a network user and its whereabouts can be monitored.

  In the $n$th time slot, the transmit powers at Alice and at the relay are denoted by $p_s[n]\in \bf{R}_+$ and $p_r[n]\in \bf{R}_+$, respectively. We suppose that the relay can store the decoded source data in the $n$th slot, and forward it in any of the remaining time slots. This imposes the \emph{information-causality} constraints, i.e.,
\begin{subequations}\label{information-causality}
\begin{align}
R_b[1]=0,\; \sum_{i=2}^{n}R_b[i]\leq\sum_{i=1}^{n-1}R_r[i], n=2,...,N,\label{IC1}\\
R_e[1]=0,\; \sum_{i=2}^{n}R_e[i]\leq\sum_{i=1}^{n-1}R_r[i], n=2,...,N,\label{IC2}
\end{align}
\end{subequations}
with $R_r[n]$, $R_b[n]$ and $R_e[n]$ being the maximum reception rates at the relay, Bob, and Eve at slot $n$, respectively.  As a basic result of (\ref{information-causality}), we set  $p_s[N]=p_r[1]=0$ without loss of optimality. Under the above setup, an achievable secrecy rate is expressed as \cite{TLSS2009}
\begin{align}
R_{s} (\{p_r[n]\},\{x[n],y[n&]\})=\sum\nolimits_{n=2}^{N}\log\left(1+p_r[n]\gamma_{rd}[n]\right)\nonumber\\
&-\sum\nolimits_{n=2}^{N}\log\left(1+p_r[n]\gamma_{re}[n]\right),
\end{align}

In this letter, our focus is on the criterion of designing $\{x[n],y[n]\}_{n=1}^{N}$, $\{p_s[n]\}_{n=1}^{N-1}$ and $\{p_r[n]\}_{n=2}^N$, under an achievable secrecy rate maximization (SRM) problem with mobility and information-causality constraints, i.e.,
\begin{subequations}\label{secrecy rate}
\begin{align}
C{_s^\star}=&\max\limits_{\begin{subarray}{c}\{x[n],y[n]\}_{n=1}^{N}\\\{p_s[n]\}{_{n=1}^{N-1}}, \{p_r[n]\}{_{n=2}^N}\end{subarray}}R_{s}(\{p_r[n]\},\{x[n],y[n]\})\\
\textrm{s.t.}\;&\sum_{i=2}^{n}\log(1+p_r[i]\gamma_{rd}[i])\leq\sum_{i=1}^{n-1}\log(1+p_s[i]\gamma_{ar}[i]),\nonumber\\
&\quad\quad\quad \quad\quad\quad\quad\quad\quad\quad\quad\quad\quad\quad n=2,...,N,\label{P1a}\\
&\sum_{i=2}^{n}\log(1+p_r[i]\gamma_{re}[i])\leq\sum_{i=1}^{n-1}\log(1+p_s[i]\gamma_{ar}[i]),\nonumber\\
&\quad\quad\quad \quad\quad\quad\quad\quad\quad\quad\quad\quad\quad\quad n=2,...,N,\label{P1b}\\
&\sum_{i=1}^{N-1}p_s[n]\leq N\bar{P}_s,\;\sum_{i=2}^{N}p_r[n]\leq N\bar{P}_r,\label{P1c}\\
&p_s[n] \ge 0, n=1,2,\cdots,N-1,\label{P1d}\\
&p_r[n] \ge 0, n=2,\cdots,N,\label{P1e}\\
&\text{(\ref{va})-(\ref{vc}) satisfied.}\label{P1f}
\end{align}
\end{subequations}
with $\bar{P}_s>0$ and $\bar{P}_r>0$ being the average power limits at Alice and the relay, respectively. It is seen in problem (\ref{secrecy rate}) that the trajectory designing and the power allocating are closely coupled with each other. This coupling makes the problem intractable and motivates us to use an alternating optimization (AO) approach \cite{QLMH2013}. 
\section{An AO Approach to the Joint Power and Trajectory Optimization}
In this section, we propose an efficient AO-based algorithm to handle the SRM problem (\ref{secrecy rate}).
Specifically, let
\begin{equation}
(\{x^m[n],y^m[n]\}_{n=1}^{N},\{p^m_{s}[n]\}_{n=1}^{N-1},\{p^m_{r}[n]\}_{n=2}^N)\nonumber
\end{equation}
 be the AO iterate at the $m$th iteration. To obtain the AO iterates for $m=1,2,...$, one can alternatingly solve two subproblems of problem (\ref{secrecy rate}), i.e., optimizing the power allocations in (\ref{secrecy rate}) with given $\{x^{m-1}[n],y^{m-1}[n]\}_{n=1}^{N}$ and optimizing the trajectory in (\ref{secrecy rate}) with fixed $\{p^{m}_s[n]\}_{n=1}^{N-1}$ and $\{p^{m}_r[n]\}_{n=2}^N$, which are defined as the \emph{power allocation} problem and the \emph{trajectory optimization} problem, respectively. We summarize the AO process in Algorithm \ref{joint_optimization}.
\begin{algorithm}
\renewcommand{\algorithmicrequire}{\textbf{Input:}}
\renewcommand\algorithmicensure {\textbf{Output:} }
\caption{An AO approach to the SRM problem (\ref{secrecy rate})}\label{joint_optimization}
\begin{algorithmic}[1]
\STATE Initialize $\{x^0[n],y^0[n]\}_{n=1}^{N}$ and set $m=1$.
\STATE \textbf{repeat} 
\STATE \quad Update $\{p^m_{s}[n]\}_{n=1}^{N-1}$ and $\{p^m_{r}[n]\}_{n=2}^N$ by solving problem (\ref{secrecy rate}) with fixed $\{x^{m-1}[n],y^{m-1}[n]\}_{n=1}^{N}$,.
\STATE \quad Update $\{x^{m}[n],y^{m}[n]\}_{n=1}^{N}$ by solving problem (\ref{secrecy rate}) with fixed $\{p^m_{s}[n]\}_{n=1}^{N-1}$ and $\{p^m_{r}[n]\}_{n=2}^N$.
\STATE \quad Update $m=m+1$.
\STATE \textbf{until} {the convergence conditions are satisfied or a maximum number of iterations has been reached.}
\end{algorithmic}
\label{algorithm2}
\end{algorithm}

Using the process above, we can tackle the SRM problem (\ref{secrecy rate}) by alternatingly solving its subproblems.
Our next endeavor is to seek solutions for the trajectory optimization problem and the power allocation problem, respectively. To begin with, we have the following lemma \cite{QW}:
\begin{lemma}
A Karush-Kuhn-Tucker (KKT) point of the power allocation problem can be obtained by exploiting the difference-of-concave (DC) algorithm.
\end{lemma}

Next, it is easy to see that the trajectory optimization problem is nonconvex. To circumvent the nonconvexity, we will utilize sequential convex programming (SCP) to tackle the trajectory optimization problem in the next section.
\begin{remark}
 In some practical scenario, the sole mission of the UAV is to assist communication so that the initial/final locations of the UAV should be optimized. In this case, the joint power and trajectory optimization problem equals to problem (\ref{secrecy rate}) without constraints (\ref{va}) and (\ref{vc}), which thus can be tackled in a similar approach.
\end{remark}
\section{A Tractable Approach to the Trajectory Optimization Problem}\label{sec:optimal-trajectory}
 In this section, we  propose an iterative algorithm based on SCP to obtain a KKT point of the trajectory optimization problem. Its basic idea is to construct successive convex approximate subproblems at successive iteration points. Notice that besides being a subproblem of (\ref{secrecy rate}), the trajectory optimization problem may also correspond to the practical scenario where Alice and the relay can only transmit with constant power because of hardware limitations.
 For ease of subsequent description, denote $\{x_l[n],y_l[n]\}_{n=1}^{N}$ as the resulting trajectory after the $l$th iteration, and $R_{r,l}[n]\triangleq\log(1+\frac{\gamma_{s}[n]}{H^2+x_l^2[n]+y_l^2[n]})$ and $R_{d,l}[n]\triangleq\log(1+\frac{\gamma_{r}[n]}{H^2+(D-x_l[n])^2+y_l^2[n]})$  
  as the reception rates at the relay and Bob at slot $n$, respectively, where $\gamma_s[n]\triangleq\frac{\beta_0 p_s[n]}{\delta^2}$ and $\gamma_r[n]\triangleq\frac{\beta_0 p_r[n]}{\delta^2}, \forall n$. Moreover, let $\{\delta_l[n],\xi_l[n]\}$ denote the trajectory variation of the UAV from the $l$th to the $(l+1)$th iteration, then we have $x_{l+1}[n]=x_l[n]+\delta_l[n]$ and $y_{l+1}[n]=y_l[n]+\xi_l[n], \forall n$.

To implement the SCP, we derive an approximate trajectory optimization problem, i.e.,
\begin{subequations}\label{convex_approx}
\begin{align}
&\max\limits_{\begin{subarray}{c}\{\delta_l[n],\xi_l[n]\}_{n=1}^{N}\\\{\epsilon[n],\tau[n]\}_{n=2}^{N}\end{subarray}}\sum_{n=2}^{N}R{_{d,l+1}^{lb}}[n]-\sum_{n=2}^{N}\log(1+\frac{\gamma_{r}[n]}{H^2+\tau[n]})\nonumber\\
 &\textrm{s.t.}\text{(\ref{P1f}) satisfied.}\label{ca_a}\\
&\sum_{i=2}^{n}\log(1+\frac{\gamma_{r}[i]}{H^2+\epsilon[i]})\leq \sum_{i=1}^{n-1}R_{r,l+1}^{lb}[i], n=2,...,N,\label{ca_b}\\
& \sum_{i=2}^{n}\log(1+\frac{\gamma_{r}[i]}{H^2+\tau[i]})\leq \sum_{i=1}^{n-1}R_{r,l+1}^{lb}[i], n=2,...,N,\label{ca_c}\\
&\epsilon[n]\ge0, n=2,...,N,\label{ca_d}\\
&\tau[n]\ge0, n=2,...,N,\label{ca_e}\\
&\zeta_{l+1}^{lb}[n]\ge \tau[n], n=2,...,N,\label{ca_f}\\
&\eta_{l+1}^{lb}[n]\ge \epsilon[n], n=2,...,N,\label{ca_g}
\end{align}
\end{subequations}
where $\{\epsilon[n]\}_{n=2}^{N}$ and $\{\tau[n]\}_{n=2}^{N}$ are slack variables introduced to simplify the problem, and $R{_{d,l+1}^{lb}}[n]$, $R{_{r,l+1}^{lb}}[n]$, $\zeta_{l+1}^{lb}[n]$, and $\eta_{l+1}^{lb}[n]$ are concave lower bounds given in (\ref{lb_a}), (\ref{lb_b}), (\ref{lb2_a}), and (\ref{lb2_b}). The development of the approximate problem (\ref{convex_approx}) is relegated to Appendix. It is not difficult to see that problem (\ref{convex_approx}) is convex; hence its optimal solution can be conveniently obtained by using a general purpose convex optimization solver, such as CVX \cite{MGSB}. Now, we can tackle the trajectory optimization problem by iteratively solving problem (\ref{convex_approx}) with updating $\{x_l[n],y_l[n]\}$. The SCP method is summarized in Algorithm \ref{algorithm1}.

\begin{algorithm}
\renewcommand{\algorithmicrequire}{\textbf{Input:}}
\renewcommand\algorithmicensure {\textbf{Output:} }
\caption{Iterative algorithm for solving the trajectory optimization problem}\label{traj_opt}
\begin{algorithmic}[1]
\STATE Initialize $\{x_0[n],y_0[n]\}_{n=1}^{N}$ and set $l=0$.
\STATE \textbf{repeat} 
\STATE \quad Invoking software CVX to obtain the optimal solution $\{\delta_l^*[n],\xi_l^*[n]\}_{n=1}^{N}$ to problem (\ref{convex_approx}).
\STATE \quad Update $x_{l+1}[n]=x_l[n]+\delta_l^*[n]$ and $y_{l+1}[n]=y_l[n]+\xi_l^*[n], \forall n$.
\STATE \quad Update $l=l+1$.
\STATE \textbf{until} {the convergence conditions are satisfied.}
\end{algorithmic}
\label{algorithm1}
\end{algorithm}
The merit of the SCP-based iterative algorithm is twofold: It only requires solving convex optimization problems, whose optimal solutions can be obtained conveniently. In addition, we have the following convergence result.
\begin{theorem}\label{kkt}
Algorithm \ref{algorithm1} is guaranteed to converge to a KKT point of the trajectory optimization problem.
\end{theorem}
\begin{IEEEproof}
To prove Theorem \ref{kkt}, we need the following facts.
\begin{enumerate}

  \item \label{property1}
  At point $\delta_l[n]=\xi_l[n]=0, \forall n$, the inequalities in (\ref{lb_a}), (\ref{lb_b}), (\ref{lb2_a}) and (\ref{lb2_b}) hold with equality.
  \item \label{property2}
  At point $\delta_l[n]=\xi_l[n]=0, \forall n$, $\nabla R_{r,l+1}[n]=\nabla R{_{r,l+1}^{lb}}[n]$, $\nabla R_{d,l+1}[n]=\nabla R{_{d,l+1}^{lb}}[n]$, $\nabla\zeta_{l+1}[n]= \nabla\zeta_{l+1}^{lb}[n]$ and $\nabla\eta_{l+1}[n]= \nabla\eta_{l+1}^{lb}[n]$, where $\nabla\mathit{f}$ denotes the gradient vector of function $\mathit{f}$.
\end{enumerate}
 Due to the page limit, the detailed proof of the above facts are omitted. By making use of Lemma \ref{lb} (cf. Appendix), (\ref{lb2_a}), (\ref{lb2_b}), as well as the above facts, Theorem \ref{kkt} is an immediate result of  \cite[Proposition 3]{AZEB}.
\end{IEEEproof}

\begin{remark}
 Algorithm \ref{joint_optimization} is now complete by utilizing the DC algorithm and Algorithm \ref{algorithm1}.
Since both the DC algorithm and the SCP only need convex problems solved, the overall complexity of Algorithm \ref{joint_optimization} is polynomial in the worst case. However, as neither the DC algorithm nor the SCP has global optimality, no optimality can be theoretically declared for Algorithm \ref{joint_optimization}.
\end{remark}

\section{Numerical Simulations}\label{sec:experiments}
In this section, numerical results are provided to demonstrate the secrecy performance of the proposed algorithms.
We consider a system where the communication bandwidth per link is $20$MHz with $5$GHz as its carrier frequency, and the noise power spectrum density is $-169$dBm/Hz.
Consequently, the reference SNR at distance $d_0 = 1$m should be $\frac{\beta_0}{\delta^2}=80$dB. We assume that the UAV flies at fixed altitude $H=100$m and its maximum speed is $\tilde V=50$m/s.
Set $E=1000$m, $S=100$m, and $D=2000$m, where $(E,S)$ denotes the x-y coordinate of Eve, and $D$ is the x-coordinate of Bob.
The maximum average transmit power at Alice and the relay are assumed to be $\bar{P}_s=\bar{P}_r=10$dBm.
\subsection{Trajectory Optimization with Fixed Power Allocation}
\begin{figure}[h]
\centering
\includegraphics[width=7cm]{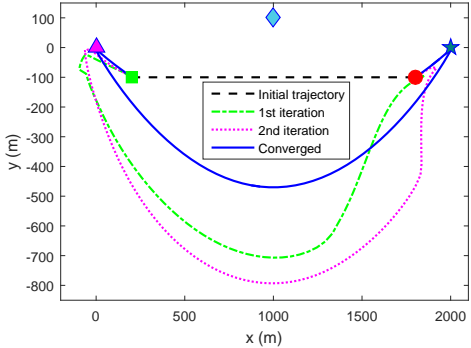}
\DeclareGraphicsExtensions.
\caption{UAV trajectory evolution by Algorithm \ref{algorithm1}. The triangle, diamond, star, quare and circle represent the Alice, Eve, Bob, and
initial and final relay locations, respectively.} \label{Traj}
\end{figure}

In this simulation, we consider fixed power allocations at the source and the relay, whereas the relay's trajectory is optimized as in Algorithm \ref{algorithm1}.
The initial and final x-y coordinates of the relay are set to be $(x_0,y_0)=(200,-100)$ and $(x_F,y_F)=(1800,-100)$, respectively, as plotted in Fig. \ref{Traj}.
Both the source and the relay are assumed to adopt the equal power allocation across different time slots. We apply Algorithm \ref{algorithm1} to successively optimize the relay
trajectory, where the relay's initial trajectory is heuristically set as directly flying from $(x_0,y_0,H)$ to $(x_F,y_F,H)$ with constant speed.

For $T=100$s, we plot in Fig. \ref{Traj} the projected relay trajectories onto the horizontal plane attained with different iterations of Algorithm \ref{algorithm1}.
As seen, the optimized trajectory does not follow the direct path. Instead, the UAV should first move toward Alice, then fly a downward curving trajectory to Bob, and lastly head to the final location. This is attributed to the fact that with sufficiently large $\tilde VT$, the relay's position can be dynamically adjusted to enhance the Alice-Bob and UAV-Bob links and impair the UAV-Eve link.
\begin{figure}[h]
\centering
\includegraphics[width=7cm]{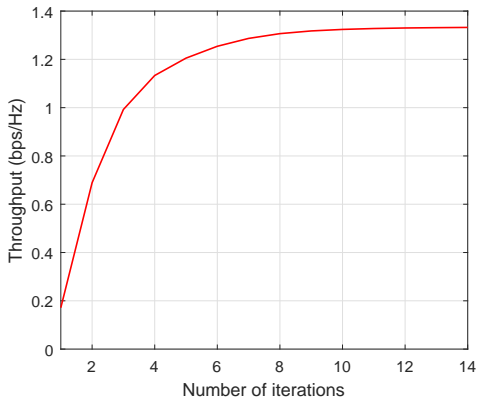}
\DeclareGraphicsExtensions.
\caption{Converge rate of Algorithm \ref{algorithm1}. } \label{Conver}
\end{figure}

Fig. \ref{Conver} plots the average secrecy rate (bps/Hz) versus the number of iterations of Algorithm \ref{algorithm1}. From the figure, one can see that Algorithm \ref{algorithm1} converges to a constant, i.e., a locally maximum throughput, in a few iterations. 
Besides, the locally maximum throughput is much higher than the throughput achieved by the initial trajectory. This observation indicates that the trajectory optimization enhances the secrecy throughput significantly, even with constant source/relay transmit power.  

\subsection{Joint Power and Trajectory Optimization}

\begin{figure}[h]
\centering
\includegraphics[width=7cm]{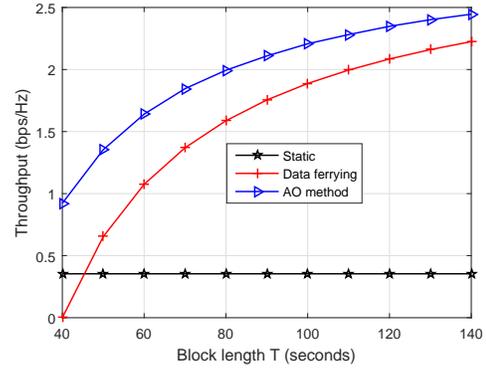}
\DeclareGraphicsExtensions.
\caption{ Average secrecy rate for mobile relaying with the proposed AO method versus static relaying and data ferrying.} \label{Compare}
\end{figure}
In this simulation, our focus is on the joint optimization of the power allocation and the relay trajectory for SRM.
In particular, we consider the scenario that the UAV has no initial or final relay locations constraint. To examine the performance of Algorithm \ref{algorithm2}, we adopt static relaying and data ferrying \cite{WZMA} as benchmarks. The maximum secrecy rate for static relaying is derived by numerically seeking the optimal fixed location at the height of $H$. For the data ferrying scheme, the UAV first hovers above Alice and receives data from Alice, flies towards Bob without any concurrent data reception/transmission, and then sends data to Bob while hovering above Bob. In Fig. \ref{Compare} we plot the average secrecy rate for mobile relaying with the proposed AO method versus static relaying and data ferrying. From Fig. \ref{Compare}, we have the following three observations: First, the proposed AO method significantly outperform the conventional static relaying scheme. Second, for small $T$, the static relaying technique even has a performance gain over the data ferrying scheme. This observation is consistent to the fact that in this case, the travel from Alice to Bob costs significant time, so that time for data loading/undloading is very limited. Third, the AO method always performs better than the data ferrying scheme.

\section{Conclusion}
In this paper, we studied the utility of mobile relaying technique in facilitating secure transmission. Specifically, we considered the joint optimization of the relay trajectory and the source/relay transmit power in maximizing the secrecy throughput, which is intractable due to the coupling of trajectory planning and power allocating.
To handle this, we proposed an AO approach, wherein the relay trajectory and the source/relay transmit power are alternatingly optimized. However, the trajectory optimization problem is a nonconvex problem, and thus is still challenging. Accordingly, we developed a SCP-based algorithm and proved that it must converge to a KKT point of the trajectory optimization problem. Our proposed algorithms only need to solve convex problems, and their overall complexities are polynomial.
In the numerical results, the provable convergence of the SCP-based algorithm was verified. Also, it was shown that the trajectory optimization can significantly improve the secrecy throughput. Moreover, the simulations demonstrated the efficacy of the proposed joint power and trajectory design in secrecy throughput enhancement.
\appendix[Development of the approximation problem (\ref{convex_approx})]\label{appendix_develop}
Our development consists of the following steps.
 The first step is to reformulate problem (\ref{secrecy rate}), with given power allocation $\{p_s[n]\}_{n=1}^{N-1}$ and $\{p_r[n]\}_{n=2}^N$, as
\begin{subequations}\label{traj_optimization}
\begin{align}
&\max\limits_{\begin{subarray}{c}\{x[n],y[n]\}_{n=1}^{N}\\\{\epsilon[n],\tau[n]\}_{n=2}^{N}\end{subarray}}\sum_{n=2}^{N}\log(1+\frac{\gamma_{r}[n]}{H^2+(D-x[n])^2+y^2[n]})-\nonumber\\
&\quad\quad\quad\quad\quad\quad\quad\quad\quad\quad\quad\quad\quad\quad\sum_{n=2}^{N}\log(1+\frac{\gamma_{r}[n]}{H^2+\tau[n]})\quad\quad\nonumber\\
&\textrm{s.t.}\text{(\ref{ca_a}), (\ref{ca_d}), and (\ref{ca_e}) satisfied.}\label{slack2_a}\\
&\sum_{i=2}^{n}\log(1+\frac{\gamma_{r}[i]}{H^2+\epsilon[i]})\leq \sum_{i=1}^{n-1}\log(1+\frac{\gamma_{s}[i]}{H^2+x^2[i]+y^2[i]}),\nonumber\\
                           &\quad\quad\quad\quad\quad\quad\quad\quad\quad\quad\quad\quad\quad\quad\quad\quad n=2,...,N,\label{slack2_b}\\
                           & \sum_{i=2}^{n}\log(1+\frac{\gamma_{r}[i]}{H^2+\tau[i]})\leq \sum_{i=1}^{n-1}\log(1+\frac{\gamma_{s}[i]}{H^2+x^2[i]+y^2[i]}),\nonumber\\
                           & \quad\quad\quad\quad\quad\quad\quad\quad\quad\quad\quad\quad\quad\quad\quad\quad n=2,...,N,\label{slack2_c}\\
                           &(E-x[n])^2+(S-y[n])^2\ge \tau[n], n=2,...,N,\label{slack2_e}\\
                           &(D-x[n])^2+y^2[n]\ge \epsilon[n], n=2,...,N.  \label{slack2_g}
\end{align}
\end{subequations}
   One can verify by contradiction that the optimal solution of (\ref{traj_optimization}) must satisfy all constraints in (\ref{slack2_e}) with equality. In addition, there always exists an optimal solution to (\ref{traj_optimization}) such that all constraints in (\ref{slack2_g}) are satisfied with equality. As a consequence, with fixed power allocation $\{p_s[n]\}_{n=1}^{N-1}$ and $\{p_r[n]\}_{n=2}^N$, problem (\ref{secrecy rate}) is equivalent to problem (\ref{traj_optimization}).
 Now, we have problem (\ref{traj_optimization}) as an equivalent trajectory optimization problem, whereas it is still nonconvex. To proceed, we will need the following lemma.
\begin{lemma}\label{lb}
For any existing relay trajectory $\{x_l[n],y_l[n]\}$ and trajectory variation $\{\delta_l[n],\xi_l[n]\}$, we have
                           \begin{subequations}
                           \begin{align}
&R_{r,l+1}[n]\ge R{_{r,l+1}^{lb}}[n]\triangleq R_{r,l}[n]-\quad\quad\quad\quad\quad\quad\quad\quad\quad\quad\quad\quad\quad\nonumber\\
&\frac{\gamma_{s}[n](\delta{_l^2}[n]+\xi{_l^2}[n]+2x_l[n]\delta_l[n]+2y_l[n]\xi_l[n])}{(d_{ar,l}^2[n]+\gamma_{s}[n])d_{ar,l}^2[n]}, \forall n\label{lb_a}\\
&R_{d,l+1}[n]\ge R{_{d,l+1}^{lb}}[n]\triangleq R{_{d,l}}[n]-\quad\quad\quad\quad\quad\quad\quad\quad\quad\quad\quad\quad\quad\nonumber\\
&\frac{\gamma_{r}[n](\delta{_l^2}[n]+\xi{_l^2}[n]+2x_l[n]\delta_l[n]+2y_l[n]\xi_l[n]-2D\delta_l)}{(d_{rd,l}^2[n]+\gamma_{r}[n])d_{rd,l}^2[n]}, \label{lb_b}\forall n
                          \end{align}
                          \end{subequations}
where $d_{ar,l}[n]=\sqrt{H^2+x_l^2[n]+y_l^2[n]}$ and $d_{rd,l}[n]=\sqrt{H^2+(D-x_l[n])^2+y_l^2[n]}$ are the resulting distances of the Alice-UAV link and the UAV-Bob link at slot n after the $l$th iteration. 
\end{lemma}
\begin{IEEEproof}
It is easy to verify that $\mathnormal{f}(z)=\log(1+\frac{\gamma}{A+z})$ is convex with regard to $z>-A$ for some constant $\gamma>0$ and $A$. By applying a first-order Taylor series expansion to $\mathnormal{f}(z)$ at point $z_0>-A$, we have the following inequality
                           \begin{align}\label{Tayler}
                           \log(1+&\frac{\gamma}{A+z})\ge\log(1+\frac{\gamma}{A+z_0})\nonumber\\
                           &-\frac{\gamma}{(A+z_0+\gamma)(A+z_0)}(z-z_0), \forall z>-A.
                           \end{align}
(\ref{lb_a}) follows from (\ref{Tayler}) by letting $\gamma=\gamma_s[n]$, $A=H^2$, $z_0=x_l^2[n]+y_l^2[n]$, and $z=x_{l+1}^2[n]+y_{l+1}^2[n]$. In a similar manner, (\ref{lb_b}) can be derived.
\end{IEEEproof}

Lemma \ref{lb} provides concave lower bounds to $\{R_{r,l+1}[n]\}$ and $\{R_{d,l+1}[n]\}$ for any existing relay trajectory $\{x_l[n],y_l[n]\}$ and trajectory variation $\{\delta_l[n],\xi_l[n]\}$. 

 Invoking Lemma \ref{lb}, our second step is to construct an approximation of problem (\ref{traj_optimization}) written as
\begin{align}\label{secondstep}
&\max\limits_{\begin{subarray}{c}\{\delta_l[n],\xi_l[n]\}_{n=1}^{N}\\\{\epsilon[n],\tau[n]\}_{n=2}^{N}\end{subarray}}\sum_{n=2}^{N}R{_{d,l+1}^{lb}}[n]-\sum_{n=2}^{N}\log(1+\frac{\gamma_{r}[n]}{H^2+\tau[n]})\nonumber\\
&\textrm{s.t.}\text{(\ref{ca_a})-(\ref{ca_e}), (\ref{slack2_e}), and (\ref{slack2_g}) satisfied.}
\end{align}
Nevertheless, problem (\ref{secondstep}) is still nonconvex due to its nonconvex constraints (\ref{slack2_e}) and (\ref{slack2_g}).
For notational convenience, denote $\zeta_{l+1}[n]\triangleq(E-x_{l+1}[n])^2+(S-y_{l+1}[n])^2$ and $\eta_{l+1}[n]\triangleq(D-x_{l+1}[n])^2+y_{l+1}^2[n]$. It is not difficult to see that both $\zeta_{l+1}[n]$ and $\eta_{l+1}[n]$ are convex. Thus, the concave lower bounds of $\zeta_{l+1}[n]$ and $\eta_{l+1}[n]$ can be easily derived by exploiting Taylor series expansions, i.e.,
\begin{subequations}\label{lb2}
\begin{align}
\zeta_{l+1}[n]\ge \zeta_{l+1}^{lb}[n]\triangleq&(E-x_l[n])^2+(S-y_l[n])^2+\nonumber\\
2(x_l[n]-&E)\delta_l[n]+2(y_l[n]-S)\xi_l[n],\forall n,\label{lb2_a}\\
\eta_{l+1}[n]\ge \eta_{l+1}^{lb}[n]\triangleq&(D-x_l[n])^2+y_l^2[n]+\nonumber\\
2(x_l[n]-&D)\delta_l[n]+2y_l[n]\xi_l[n],\forall n.\label{lb2_b}
\end{align}
\end{subequations}

Our third step is to reformulate problem (\ref{secondstep}) by approximating $\{\zeta_{l+1}[n]\}$ and $\{\eta_{l+1}[n]\}$ in  (\ref{slack2_e}) and (\ref{slack2_g}) with their concave lower bounds $\{\zeta_{l+1}^{lb}[n]\}$ and $\{\eta_{l+1}^{lb}[n]\}$, respectively. Then, the resulting approximate trajectory optimization problem, i.e., problem (\ref{convex_approx}), is convex.

\bibliographystyle{IEEEtran}
\bibliography{newbib}
\end{document}